\documentstyle[12pt]{article} \input epsf

\input epsf

\begin{document}

\renewcommand{\thesection}{\arabic{section}.} 
\renewcommand{\theequation}{\thesection \arabic{equation}}
\newcommand{\scs}{\setcounter{equation}{0} \setcounter{section}}
\def\req#1{(\ref{#1})}
\newcommand{\be}{\begin{equation}} \newcommand{\ee}{\end{equation}} 
\newcommand{\ba}{\begin{eqnarray}} \newcommand{\ea}{\end{eqnarray}} 
\newcommand{\la}{\label} \newcommand{\nb}{\normalsize\bf} 
\newcommand{\lb}{\large\bf} \newcommand{\vol}{\hbox{Vol}}
\newcommand{\bb} {\bibitem} \newcommand{\np} {{\it Nucl. Phys. }} 
\newcommand{\pl} {{\it Phys. Lett. }} 
\newcommand{\pr} {{\it Phys. Rev. }} \newcommand{\mpl} {{\it Mod. Phys. Lett. }}
\newcommand{\sg}{{\sqrt g}} \newcommand{\sqhat}{{\sqrt{\hat g}}}
\newcommand{\sqphi}{{\sqrt{\hat g}} e^\phi} 
\newcommand{\sqalpha}{{\sqrt{\hat g}}e^{\alpha\phi}}
\newcommand{\tp}{\cos px\ e^{(p-{\sqrt2})\phi}} \newcommand{\stwo}{{\sqrt2}}
\newcommand{\tr}{\hbox{tr}}

\begin{titlepage}
\renewcommand{\thefootnote}{\fnsymbol{footnote}}

\hfill CERN-TH/99--405

\hfill hep-th/9912155

\vspace{.4truein}
\begin{center}
 {\LARGE AdS--Flows and Weyl Gravity}
 \end{center}
\vspace{.7truein}

 \begin{center}

 Christof Schmidhuber\footnote{schmidhu@mail.cern.ch}

 \vskip5mm

 {\it CERN, Theory Division, 1211 Gen\`eve 23, Switzerland}

 \end{center}

\vspace{1truein}
\begin{abstract}
\noindent
An analogy is noted between the RG flow equations  
in 4--dimensional gauge theory, as derived from the AdS/CFT
correspondence, and the RG flow equations
in 4--dimensional field theory coupled to a 
particular limit of Weyl supergravity. This suggests
a possible theory of dynamical 3-branes 
with fluctuating 4--dimensional conformal factor.
The argument involves a map from
flows in 4-dimensional gauge theories
to flows in a class of 2-dimensional sigma models.

\vfill
\noindent
 CERN-TH/99--405

\noindent
December 99

\end{abstract}
 \renewcommand{\thefootnote}{\arabic{footnote}}
 \setcounter{footnote}{0}
\end{titlepage}

{}\noindent
\subsection*{1. Introduction} \scs{1} 

According to recent conjectures, 
${\cal N}=4$ supersymmetric $SU(N)$ 
Yang Mills theory with `t Hooft coupling
$$\lambda\ =\ g_{YM}^2N\ ,$$ 
as well as various other conformally invariant $4d$ gauge theories
have strong coupling descriptions in terms of type IIB superstring theory
on $AdS_5$ times various Einstein manifolds \cite{mal, kle, kgp, wit1, klewit, gub, oz}.
String loop corrections are proportional to ${1\over N^2}$,
while $\alpha'$-corrections are proportional to
${1\over\sqrt\lambda}$.
In particular, the strong-coupling limit $\lambda\rightarrow\infty$
can be investigated in the supergravity approximation.

This
dual description has been used to study RG flows in $4d$ gauge theory
in the large-N, large-$\lambda$ limit,
based on the interpretation of the radial
coordinate of $AdS_5$ as the scale of the $4d$ theory \cite{itz,porr,war,oth,ver3}. The
type IIB supergravity equations of motion then become RG flow equations.

Here it is first demonstrated how - at least in 
the vicinity of fixed points -
these $4d$ flows can be related to flows in $2d$ sigma models
with $5d$ target space and Ramond-Ramond backgrounds. 
Fixed points are mapped to fixed points,
and c-functions, beta functions and phase diagrams
of the $4d$ and $2d$ theories are related to each other.
This puts the
previous supergravity results into a form
in which they might be extendable to all orders in $\alpha'$.

It is then noted that the $4d$ flows look very much
like flows in $4d$ field theories coupled to ``conformally self-dual
$4d$ gravity'' \cite{ft3,ant,schm4d}, where the dynamics of the conformal factor originates from
the $4d$ conformal anomaly.
It is speculated that this suggests that the world-brane theory
of $N$ D-branes is really
a theory that contains this version of $4d$ supergravity; and
that this suggests a theory of 
dynamical 3-branes that might be analogous to the Polyakov formulation of
dynamical strings.

\subsection*{2. General setup}\scs{2}

We consider type IIB string theory in an $AdS_5\times E^5$ background.
$E^5$
is an Einstein manifold, so its Ricci tensor is $R_{mn}={\Lambda} g_{mn}$
with some $\Lambda$ and $m,n\in\{5,6,7,8,9\}$.
Let us parametrize the Einstein manifold by coordinates $\theta^m$ 
and denote by $x_{||}$ the coordinates $x^\mu$, $\mu\in\{0,1,2,3\}$,
 of the $4d$ space
parallel to the boundary of $AdS_5$. The radial coordinate of $AdS_5$
will be called $\phi$.
The $AdS_5\times E^5$ metric is then
\ba ds^2=d\phi^2+e^{2\alpha(\phi)}dx_{||}^2+L^2\hat g_{mn}d\theta^md\theta^n,
\la{fredy}\ea 
where
$\hat g$ is defined such that the volume of $E^5$ as
measured in the metric $\hat g$ is one; by a proper choice of coordinates,
$\hat g$ can also locally be made to have unit determinant:
$$\hbox{Vol}\ \hat E^5\ =\ 1\ \ \ ,\ \ \ \det\ \hat g\ =\ 1\ .$$
$L$ is a parameter that is related to $\Lambda$ by
$$\Lambda\ =\ {\lambda\over L^2}\ \ ,\ \ \ \hbox{where}\ \ \ \hat R_{mn}=\lambda\hat g_{mn}\ .$$
With these definitions, we find that $E^5$ and $AdS_5$ have the same curvature scalar up to a sign
if $5\Lambda=16\dot\alpha^2$ (``dot'' means ``$d/d\phi$''), i.e.
\ba
\alpha(\phi) &=&-{q\over4}\ {\phi\over L}\ \ \ \hbox{with}\ \ \ \ {q}\ \equiv\ 
{\sqrt {5\lambda}}\ .\la{freby}\ea
The sign is such that the boundary of $AdS_5$ is at $\phi\rightarrow-\infty$.
$\hat g$, $L\equiv e^\beta$ and the dilaton $\Phi$ are independent of $\phi$:
\ba
\hat g&=&\hat g_0\ =\ const.\\
\beta&=&\beta_0\ =\ const.\\
\Phi&=&\Phi_0\ =\ const.\ea
All other fields are zero, except that there
are also $N$ units of electric and magnetic Ramond-Ramond 5-form flux:
$$\oint_{E^5} F^{(5)}\ =\ \oint_{E^5} *F^{(5)}\ \sim\ N\ ,$$
such that
$$
-\ g^{00}g^{11}g^{22}g^{33}g^{44}F_{01234}F_{01234}\ =\   
g^{55}g^{66}g^{77}g^{88}g^{99}F_{56789}F_{56789}\ \sim\ {N^2\over L^{10}}\ .
$$

Now, the
target space fields of type IIB superstring theory appear
as coupling constants {\it both} in the $4d$ Yang-Mills theory
and in the $2d$ sigma model that lives on the string world-sheet.
This is just the general statement (which does not depend on
having an AdS--background) that type IIB string fields can be
coupled both to D3--branes and to fundamental 1--branes.
In the $4d$ Lagrangean the string fields appear as follows:
\ba {\cal L}_{YM} &\sim& e^{-\Phi}\ \tr\ F^2\ \\
&+&\chi\ \tr\ F\wedge F\ \ \\
&+& \bar g_{ab}\ \tr\ \partial\phi^a \partial\phi^b\\
 &+& A^{(4)}_{abcd}\ \tr\ \partial \phi^a\wedge\partial \phi^b
\wedge\partial \phi^c\wedge\partial \phi^d\ +\ ...\ea
Here, $\chi$ and $A^{(4)}$ are the Ramond-Ramond 0-form and 4-form, and
$\phi^a$ are the scalar fields of the SYM theory, where $ a,b\in\{1,...,6\}$.
We have omitted the two-form gauge fields, but they can be included.
So we have the well-known relations
\ba  e^{\Phi}&=&g^2_{YM}\\
\chi&=&\theta-\hbox{angle}\ ,\ea
etc. Furthermore, $\bar g$ is directly related to $\hat g$. E.g., in the case $E^5=S^5$,
$\bar g=\delta$ and, using spherical coordinates $(r,\theta^m)$
in $\phi^a$--space, we have
$$ \bar g_{ab}\ d\phi^a d\phi^b\ \sim\ dr^2\ +\ r^2\hat g_{mn}(\theta)d\theta^m d\theta^n\ .$$

$\alpha(\phi)$, on the other hand, is not a coupling constant
of the Yang-Mills theory. It will be regarded
as an auxiliary parameter below and eliminated by its equations of motion.
$L$ is also not an independent Yang-Mills coupling constant, and will be
regarded as an auxiliary parameter, too --
its equation of motion relates it, at fixed points, to the
`t Hooft coupling:
$$L^4\ \sim\ {g^2_{YM}N}\ .$$

As mentioned, the same type IIB string fields also
appear as coupling constants in the $2d$ Lagrangean of the world-sheet sigma model, but of course differently:
\ba
{\cal L}_{\sigma}&\sim&R^{(2)}\Phi\\
&+&(\partial \phi)^2\\
&+&L^2\hat g_{mn}\partial \theta^m\partial\theta^n\\
&+&e^{2\alpha}\partial x_\mu\partial x^\mu\\
&+& \hbox{RR-backgrounds}\ea
Now, this sigma model contains 
more coupling constants than the Yang-Mills theory:
it contains the auxiliary parameters $\alpha(\phi)$ and $L$,
and it also allows in principle for a non--flat $4d$ metric $g_{\mu\nu}$.
It is therefore natural to instead consider a ``reduced''
 sigma model that accounts {\it only} 
for the YM coupling constants - namely
the sigma--model with only $5d$ internal target space,
parametrized by $\theta^m$, rather than the full $10d$ target space,
and without the $L^2$ factor in the
embedding space metric:
$$\hat{\cal L}_{\sigma}\ \sim\ R^{(2)}\Phi \ +\ 
\hat g_{mn}\partial \theta^m\partial\theta^n\ +\ \hbox{RR-backgrounds on $E^5$}.$$
Then fixed points of the $4d$ gauge theory
(with only $\alpha$, but not
the dilaton and the metric of the Einstein manifold  
changing as a function of $\phi$)
correspond to fixed points of this $2d$ sigma model.
We want to suggest how furthermore, at least in supergravity approximation,
$4d$ c-functions are mapped to $2d$ c-functions, based on \cite{war},
and how the whole flow in the $4d$ gauge theory is mapped onto a 
flow in the $2d$ sigma model for the internal compact space.

This relation between the RG flow in this sigma model with $5d$ target space
and the flow in $4d$ Yang--Mills theory will be based on the interpretation of $\phi$
in terms of the scale of the Yang--Mills theory \cite{suss,peet}: for
the AdS metric in (\ref{fredy},\ref{freby}), we can absorb overall  scale transformations on the brane
$$x_{||}\ \rightarrow\ x_{||}\ e^\tau$$
in shifts 
$$\phi\ \rightarrow\ \phi\ +\ {4\over q}\ L\tau\ .$$
In this sense, $\phi$ is ``RG time'' (the UV end $\tau=-\infty$ is at the AdS boundary at $\phi=-\infty$), and the $\phi$-dependence of the string fields
$\hat g,\Phi, \chi, A^{(4)},...$ describes how the coupling constants
``run'' under scale transformations.
RG trajectories in the gauge theory are thus viewed as time-dependent classical solutions
$\{\hat g(\phi),\Phi(\phi), \chi(\phi), A^{(4)}(\phi),...\}$ of string theory
\cite{porr,war}. \footnote{This is quite 
analogous to the situation in the case of the RG flow in $2d$ field theory
coupled to gravity, where RG trajectories are also classical solutions of 
string theory \cite{flow}. A difference is that in that case
$\phi$ represents the $2d$ scale, while here it represents the $4d$ scale.}

If $\hat g,\Phi,\chi,A^{(4)},...$ (and, as a consequence, $L$ and $\dot\alpha(\phi)$)
are constant, we have a scale invariant theory or RG fixed point
-- {\it both} in the $4d$ and the $2d$ field theories.
We will study the flow in the 
vicinity of such fixed points:
\ba
\hat g_{mn}&=&\hat g_{mn}^0+\delta\hat g_{mn}(\phi,\theta)\\
\Phi&=&\Phi_0+\delta\Phi(\phi,\theta)\\
\chi&=&\delta\chi(\phi,\theta),
\ea
and so on. 
This is the most general ansatz consistent
with Poincar\'e invariance on the brane (we also make the local gauge choice
$g_{\phi\mu}=g_{\phi m}=0$).
We collectively denote (the modes of) the string field variations --
i.e. the coupling constants of both the $4d$ and $2d$ field theory -- by
$\vec\lambda$:
$$\vec\lambda(\phi)\ \equiv\ \{\delta\hat g_{mn},
\delta\Phi,\delta\chi,...\}\ .$$
and assume that all the coupling constants $\vec\lambda$ are small.

As the other target space fields vary, $L=e^{2\beta}$ and $\alpha$ will also adjust:
$$\beta\ =\ \beta_0+\delta\beta\ ,\ \ \ \alpha\ =\ -\ {q\over4}\ {\phi\over L}\ +\ \delta\alpha\ .$$
Then the scale $e^\tau(\phi)$ is (by holography)
$$(e^\tau)'(\phi)\ =\ e^{-\alpha(\phi)}.$$
The auxiliary parameters $\alpha,\beta$ 
will be eliminated below by solving their equations of motion
in the vicinity of fixed points.

\subsection*{3. $4d$ flows from $2d$ flows}\scs{3}

The relation we propose between the $4d$ and $2d$ flows is derived in
appendix A in the classical supergravity approximation following
the method used in \cite{st}. This approximation is valid
in the limit of large $N$ and large `t Hooft coupling $g_{YM}^2N$. The
result is the following.
Write the perturbed gauge theory Lagrangean as
$${\cal L}_4\ =\ {\cal L}_4^{0}\ +\ \vec\lambda\cdot\vec O_4\ ,$$
where ${\cal L}_4^{0}$ corresponds to the $4d$ CFT,
whose central charge we call $c_4$.\footnote{I.e. the conformal anomaly coefficients are $c_4=c=a$.}
Write the perturbed action of the $2d$ sigma model with $5d$ target space 
parametrized by $\theta^m$, and target space metric $\hat g$ (and not $g$!), as:
$${\cal L}_2\ =\ {\cal L}_2^{0}\ +\ \vec\lambda\cdot\vec O_2\ ,$$
where ${\cal L}_2^{0}$ corresponds to the $2d$ CFT with central charge $c_2$.
Call the beta functions in the $2d$ sigma model $\vec\beta^{(5)}$, so the
$2d$ flow obeys
\ba\dot\lambda_i\ =\ \beta^{(5)}_i\ =\ \Delta_i^j\lambda_j\ +\ c_i^{jk}\lambda_j\lambda_k\ +\ ...\ ,\la{david}\ea
where $\Delta_i^j$ are the dimensions of the $2d$ field theory,
$c_i^{jk}$ are the OPE coefficients, and more generally,
the beta functions are gradients of the $2d$
$c$-function $c_2(\vec\lambda)$.
Then the flow of the same coupling constants in the $4d$ gauge theory obeys
in the vicinity of fixed points, i.e.
to linear order\footnote{there are
nonuniversal corrections of order $\dot\lambda^2$ to the LHS;
also, at next order, the RHS of eqn. (\ref{max}) contains a term proportional to $\dot\lambda^2$ (see (\ref{ganz})).}
in $\dot\lambda$, $\lambda$:
(but possibly to all orders in $\alpha'$)
\begin{eqnarray}
  {q^2\over16}(\ddot{\vec\lambda}+{4}\dot{\vec\lambda}) &=&
      \vec\beta^{(5)}\la{marius}\\ \hbox{with}\ \ 
 q^2 &=& {5-c_2\over3}\ +\ \hbox{h.o.}\la{max}
\end{eqnarray}
The statement is that on the RHS of (\ref{marius}) we have the {\it same} beta functions (\ref{david}) as in the $2d$ sigma model, only the LHS changes in going from the $2d$ sigma model to the $4d$ gauge theory. 
This implies that fixed points are mapped to fixed points, and
the structure of fixed points and flows between them in $4d$ gauge theory
seems to be at least as rich as in $2d$ field theory.
Moreover, there might be a natural way in which the $c$--function
of $2d$ field theory induces a $c$--function of $4d$ gauge theory, as discussed
in point 4 below.

In the following we will occasionally redefine
$\tau\rightarrow\tilde\tau={4\over q}\tau$. Then the equation
becomes
\ba
\ddot{\vec\lambda}\ +\ q\dot{\vec\lambda}\ =\ \vec\beta^{(5)}\ ,\ \ \ \ 
\hbox{with}\ \ \ \ q^2\ =\ {5-c_2\over3}\ .\la{stern}\ea
To compare with some of the literature \cite{porr}, \cite{war}, 
note that the same equations also apply to
gauged supergravity.
In this case the coupling constants $\vec\lambda$
correspond to the 42 scalar fields $\Phi^J$ that
represent deformations of the metric, B-fields, $\chi$ and $\Phi$,
 and the $c$-function $c_2(\vec\lambda)-5$
is represented by the
the potential $v(\Phi)={1\over L^2}V(\Phi)$.
As in \cite{porr,war}, in the supergravity approximation it can be seen
   from the appendix (cf. eqn. (\ref{marcel}))
   that it is consistent to set the dilaton $\Phi$ constant.
   So the Yang--Mills coupling constant 
   $g_{YM}=e^{\Phi\over2}$ does not run: it remains large and thus
it is consistent to use the supergravity approximation.

In the case of  supersymmetric flows describing BPS-saturated
kink solitons of gauged supergravity \cite{war}, the second-order
flow equations become standard first-order flow equations
(the Bogomoln'yi equations). Moreover, the beta functions
become independent of $g_{YM}$ and can be interpolated from strong 
coupling to the perturbative regime.

Even without supersymmetry, it
is possible to define modified beta functions 
$${{\beta^{(4)}}^i}\ =\ {\Delta^{(4)}}^i_j\lambda^j\ 
+\ {c^{(4)}}^i_{jk}\lambda^j\lambda^k\ ,$$
such that the $4d$ flow becomes first--order: $\dot\lambda^i={{\beta}^{(4)}}^i$.
Plugging the first--order equations into the second order equations
(\ref{marius}) 
yields the modified beta function coefficients order by order (see below and
\cite{schmTop}). 

The above statements are well-known in the
supergravity approximation. One reason for rewriting them in the
 form of a relation between $4d$ gauge theory and
$2d$ field theory (rather than supergravity)
is to motivate the comments in the next section. 
Another reason is that now
it makes sense to ask 
whether relation (\ref{marius},\ref{max})
might hold, in the vicinity of fixed points, beyond the supergravity approximation. Before commenting on this, 
however, let us summarize what the relation is good for:
\begin{enumerate}
\item {\it Dimensions:}
Keeping only the linear part of the $2d$ beta functions
$\beta_i=\Delta^{(2)}_i\lambda_i$, the dimensions $\Delta_i^{(4)}$ of the
 $4d$ operators obey
$$\Delta^{(4)}(\Delta^{(4)}+4)\ =\ {16\over q^2}{\Delta^{(2)}_i}\ .$$
We have to drop one of the two ``dressings''. 
In supergravity approximation, where
$$q^2\ =\ {1\over3}(c_2-5)\ =\ \hat R\ =\ 5\lambda\ ,$$
this relation takes on a form that is well-known \cite{kgp,wit1}.
\item {\it OPE coefficients:}
the quadratic beta function coefficients $c^{(2)}_{ijk}$ are universal if $\Delta^{(2)}_i$ is small
(of order $\lambda$).
To relate the OPE coefficients in two and four dimensions,
we assume that $\Delta^{(2)}_i=0$ and
make the ansatz that the $4d$ flow $\dot\lambda$ in (\ref{marius})
also obeys a first-order differential equation, similarly as in \cite{schmTop}:
$$\dot \lambda \ = \ 
c_{ijk}^{(4)}\lambda^j\lambda^k\ +\ ...$$
Plugging this into (\ref{marius}) and expandig yields:
$$ c^{(4)}_{ijk}\ =\ {4\over q^2}\ c^{(2)}_{ijk} \ .$$
\item
{\it Phase diagrams:}
since fixed points are mapped to fixed points and relevant, marginal
and irrelevant directions of the flow near fixed points
are again mapped onto
relevant, marginal
and irrelevant directions, the phase diagrams of the $4d$ and $2d$
theories should also agree. 
\item
{\it $c$-functions:}
The $2d$ $c$-function $ c_2(\vec\lambda)$
always decreases along the $2d$ flow,
as the beta functions are its gradients.
But (\ref{marius}) describes the damped motion of a particle in
the potential $c_2(\vec\lambda)$,
so at least near the fixed point $c_2(\vec\lambda)$  also decreases
along the $4d$ flow. 
In this sense,
the $c$-function of $2d$ field theory might also give rise to a $c$-function of 
$4d$ gauge theory.
Away from the fixed point it seems hard to make a clear statement.

More generally, a good $4d$ c-function $c_4(\vec\lambda)$
might be some function of the $2d$ c-function $c_2(\vec\lambda)$.
To see precisely {\it what} function, note that it
has been argued in \cite{war} within the supergravity approximation
that $(\dot\alpha)^{-{3}}$ ($\dot\alpha$ is defined in section 2)
 is proportional to to the $4d$
conformal anomaly at fixed points and might be taken as a $4d$ $c$-function.
Since $\dot\alpha^2\propto(\bar c-5)$ at fixed points
(see appendix A), it
is tempting to conjecture 
a possible generalization of this statement to
higher orders in $\alpha'$: that the 
$4d$ $c$--function follows from the $2d$ $c$-function by
$$[c_4(\vec\lambda)-k]\ \sim\ -[c_2(\vec\lambda)-5]^{-{3\over2}}\ ,\ $$
where we have allowed for some constant $k$. The relative minus sign
is needed to make $c_4$ {\it de}crease, rather than {\it in}crease along
the flow.

\end{enumerate}

Could the relations (\ref{marius},\ref{max}) be, in the vicinity of fixed points,
 exact in $\alpha'$?
One possibility to adress this question would be to include $\alpha'$-corrections
to the string effective action and check whether they 
are of higher order in $|\vec\lambda|$, i.e. vanish in the vicinity
of fixed points. 

Another possibility is suggested by
comparing with an analogous situation that arises in quite a different context:
suppose we perturb a $2d$ CFT:
$${\cal L}\ =\ {\cal L}\ +\ \lambda^iO_i$$
Then the coupling constants will typically flow under scale transformations:
$$\dot{\vec\lambda}\ =\ \vec\beta\ .$$
How is this flow modified if the $2d$ field theory is coupled to gravity? The answer is (see the appendix of \cite{schmHab} for a review):
$${\alpha^2\over4}(\ddot{\vec\lambda}\ +\ Q\ \dot{\vec\lambda})\ =\ \vec\beta\ \ \ \ \hbox{with}\ \ \ Q^2={D_{crit}-c\over3}\ $$
and some constant $\alpha$.
Now in $2d$ gravity we can prove that these equations are exact in $\alpha'$ in the vicinity of
the fixed point (i.e. to order $|\vec\lambda|$), and hold
not just in the supergravity approximation.
This works as follows \cite{ddk}:
The CFT coupled to gravity is described by a CFT with one more field $\phi$, the Liouville mode.
At the fixed point, the two CFT's decouple:
$$S_{CFT}\ +\ \int d^2\xi\ (\partial\phi^2+Q\hat R\phi)\ .$$
Using OPE's, we know the exact central charge $c$ and dimension of 
the operator $e^{\gamma\phi}$ in the second CFT:
$$c=1+3Q^2\ \ \ \ ,\ \ \ \ \ 
\hbox{dim}(e^{\gamma\phi})\ \propto\ \gamma(\gamma+Q).$$
Conformal invariance of
the perturbed theory then implies the two flow equations above \cite{flow},
defining
$\tau={\alpha\over2}\phi$.
In the case of $4d$ SYM, it is of course much harder to check whether the flow equations are exact in $\alpha'$ because of the 
Ramond-Ramond backgrounds.\footnote{See \cite{berk} for attempts
to deal with Ramond-Ramond backgrounds. One should first
fully understand the simpler case of the $2d$
conformal field theories that are dual to string theory on $AdS_3$ times some
compact manifolds, where there are no RR backgrounds; see, e.g., \cite{sei} 
in this context.}
It might be possible, though, to turn the Ramond--Ramond backgrounds
in the $\phi-x^\mu-$sector (which is the analog of the above $\phi$--sector)
into Neveu-Schwarz backgrounds by duality transformations.
We must leave this subject
to future work.
\vskip4cm

\subsection*{4. Dynamical $4d$ conformal factor ?}\scs{4}

Let us now come to a key point of this paper.
Second--order flow equations like (\ref{marius}) seem
to be typical for RG flows in theories with a {\it fluctuating} scale, 
as in the case of $2d$ gravity just mentioned.
In fact, the equation (\ref{stern}),
\ba\ddot{\vec\lambda}\ +\ q\ \dot{\vec\lambda}\ \propto\ \vec\beta\ ,\ \ \ 
\ q^2\ \propto\ \hbox{conformal anomaly}\la{tim}\ea
looks precisely like the flow equation in $4d$ field
theory coupled to a particularly interesting limit of Weyl supergravity:
``conformally self--dual gravity''.
By ``conformally self--dual gravity'' we mean $4d$ gravity with action
\ba\rho\int d^4x{\sqrt g}\ W^2\ \ \ \ \ \hbox{in the limit}\ \ 
\ \ \rho\ 
\rightarrow\ \infty\ ,\la{label}\ea
where $W^2$ is the Weyl${}^2$ term. The cosmological constant,
and the coefficients of the $R$ and $R^2$ terms are assumed to be zero.
This theory is not quite renormalizable, because there is also
a conformal anomly term; but in the limit $\rho\rightarrow\infty$
the anomaly term can nicely be dealt with, as will be discussed below.

However, one point must be adressed before going any further.
There is a well-known problem
common to all fourth-order derivative actions, like the Weyl${}^2$ action:
we can rewrite them in terms of new fields with two derivatives
only, but some of them will have the wrong sign in the kinetic term.
With Minkowskian signature, this leads to perturbative non-unitarity.
For this reason, we can only consider Euclidean signature in 
this section.

Below it will be explained why this limit of Weyl gravity leads to 
 flow equations of the type (\ref{tim}). 
But before, let us
give another independent argument that indicates that
conformally self--dual gravity
might indeed arise as part of the world--volume theory.\footnote{For 
another (different) discussion of the role of Weyl gravity
in the context of the AdS/CFT correspondence, see
\cite{liu}.}

\vskip2mm\noindent
$\underline{\hbox{Kaluza-Klein reduction}}$

\noindent
Why might (\ref{label}) arise as part of the world--brane theory?
Let us start with the dual formulation of the D3--world--brane theory
in terms of 5-dimensional
gauged supergravity in an $AdS_5$ background.
Suppose we further ``Kaluza-Klein reduce'' this $5d$ gravity theory
along $z=\exp\{{q\over4}{\phi\over L}\}$
onto the $4d$ boundary of $AdS_5$ (similarly as in \cite{rs}). 
By this we mean that all fields except for the
``warp factor'' in the line element 
${1\over z^2}dx_{||}^2$ are assumed to
be independent of the 5th coordinate $z$, and $z$ is integrated
over. (In standard Kaluza-Klein reduction there would be no warp factor
and $z$ would be some circular coordinate.) 
Now, as in standard Kaluza-Klein reduction, the 
$5d$ Hilbert-Einstein action
induces a $4d$ Hilbert-Einstein action.
Its coefficient
(the inverse $4d$ Newton constant) diverges (we neglect factors of $L$):
$${1\over\kappa_5^2}\int_\epsilon dz {\sqrt{g^{(5)}}}R^{(5)}
\ \ \rightarrow\ \ {1\over\kappa_4^2}{\sqrt{g^{(4)}}}R^{(4)}\ \ \ \ \hbox{with}
\ \ \ \ 
{1\over\kappa_4^2}\ \sim\ {1\over\kappa_5^2}\int^\infty_{\epsilon} 
{dz\over z^3}\ \sim\ {1\over\epsilon^2}\ ,$$
where we have cut off $AdS_5$
at some small distance $z=\epsilon$ from the boundary.
This divergence is
powerlike and not universal,
in the sense that it depends
on precisely {\it how} we cut of $AdS_5$ at its boundary. 
There is however a universal $\log\epsilon$ divergence of the form \cite{sken}
$$\rho\int d^4x{\sqrt g}(\ W^2\ +\ G)\ \ 
,\ \ \ \ \hbox{with}\ \ \ \rho\ =\ c\ \log\epsilon\ 
\rightarrow\ \infty\ ,$$
where  the number $c$
is proportional to the $4d$ conformal anomaly
and $G$ is the Gauss--Bonnet density, whose integral is a topological
invariant (the Euler character). 
So if we assume that we can pick a
``renormalization scheme'' (i.e. a way to regularize the divergences
at the boundary) in which the non-universal Hilbert-Einstein (and
cosmological) term vanishes, then what remains is -- up to a topological
invariant --
indeed the ``conformally self-dual gravity'' action.

Precisely the bosonic sector of this supergravity theory in the limit
$\rho\rightarrow\infty$
has previously been studied by Fradkin and Tseytlin
\cite{ft3} and as a theory closely analogous to 
$2d$ gravity by Antoniadis and Mottola \cite{ant} and 
by the author \cite{schm4d}.
It is a renormalizable theory
whose nice properties are summarized in appendix B;
the main points will be recalled below. 

Is this version of $4d$ gravity dynamical here, or does it correspond to
a fixed background? There is
a standard argument that says that there is no
dynamical $4d$ gravity on the $AdS$--boundary (see e.g. \cite{oz}). It asserts
that the norm of those states of the $5d$ theory that would
correspond to $4d$ gravitons diverges, so these states are not normalizable. 
However, this argument may not apply here, because
 this divergence of the norm
is -- like the inverse Newton constant -- not universal. 
The divergence may also vanish in a scheme
in which the the inverse Newton constant
vanishes, leaving us with a finite norm state. 

Let us therefore try out the hypothesis that
dynamical conformally self--dual $4d$ gravity is indeed present on the 3--brane,
and show that this could reproduce at least qualitatively the form
 of the flow equation (\ref{tim}).

\vskip2cm\noindent
$\underline{\hbox{Conformally self-dual gravity}}$

\noindent
The following is briefly reviewed in appendix B:
up to topological invariants the $4d$ gravity action to be studied is
actually \cite{schm4d}
$$\rho\int d^4x\ {\sqrt g}\ W_+^2\ \ ,\ \ \ \ \rho\rightarrow\infty\ .$$
$W_+$ is the self-dual part of the Weyl tensor. The limit $\rho\rightarrow\infty$
restricts the metric to be
``conformally self-dual'': $W_+=0$ (see appendix).
The path integral over metrics then reduces,
in analogy with $2d$ gravity, to a path integral
over the conformal factor $\phi$ and an integral over the moduli space of
conformally self-dual metrics.

In $2d$ gravity, the dynamics for the conformal factor arises
from the nonlocal conformal anomaly term in the effective action.
In conformal gauge the anomaly term becomes local; it
consists of a kinetic term for $\phi$ plus a background charge term
 (neglecting the Liouville potential)
\ba\phi\Box\phi\ -\ 
QR\phi\ \ \ \hbox{with}\ \ \ Q^2={c-25\over3}\ .\la{else}\ea
$Q$ is determined to make the total conformal anomaly vanish.
In our $4d$ theory, everything is almost completely analogous \cite{ft2,ant}
 (see appendix B). The nonlocal anomaly term becomes a local term for $\phi$,
 if evaluated for conformally self-dual metrics (which are the only metrics
that survive the limit $\rho\rightarrow\infty$). The 
relevant part of the induced Lagrangean for $\phi$ is
$$
 \phi \Box^2 \phi\ -\ {1\over2}Q  G\phi\ \ \ \hbox{with}
\ \ \ Q^2={\tilde b+1538\over90}\ $$
with Gauss-Bonnet density $G$ and $\tilde b=-360b$ where the 
$4d$ conformal anomaly coefficients are called $a,b$: essentially,
$$<T^\mu _\mu> \ =\ {1\over{16\pi^2}}\ (aW^2 + bG) \ $$
(assuming $g$ is an Einstein metric).
Some coefficients $a$ (which plays a minor role here)
 and $b$ are given in the appendix.

Due to the $\Box^2$ kinetic term for $\phi$, its propagator is logarithmic
and therefore the exponential
$e^{2\alpha\phi}$ has definite scaling dimension \cite{ant,schm4d}
$$\dim(e^{\alpha\phi})=-\alpha(\alpha+Q).$$
As in $2d$, scaling operators $\Phi_i$ of the matter theory
with dimension $\Delta_i$ are gravitationally dressed:
$\Phi_i\rightarrow e^{\gamma_i \phi} \Phi_i$. Scale ivariance determines
the dressed scaling dimension $\gamma_i$:
$$\Delta_i-\gamma_i(\gamma_i+Q)=4$$
Now, perturbing the CFT by operators $\lambda^ie^{\gamma_i \phi}\Phi_i$,
defining $\lambda^i(\phi)=\lambda^ie^{\gamma_i \phi}$, 
multiplying the above equation by $\lambda^i(\phi)$, replacing 
$\gamma_i\rightarrow\partial_\phi$ and identifying
$\beta_i\ =\ (\Delta_i-4)\lambda_i\ +\ O(\lambda^2)$
yields the following flow equations:
\ba\ddot{\vec\lambda}\ +\ Q\dot{\vec\lambda}\ \propto\vec\beta\ ,
\ \ \ \ Q^2\ \propto\ \hbox{conformal anomaly}\ .\la{tom}\ea
These are RG flow equations in view of the interpretation
of $\phi$ as the $4d$ conformal factor -- so overall $4d$
 scale transformations correspond to constant shifts of $\phi$.

These flow equations indeed have the same form as (\ref{tim}).
It is tempting to interpret this as additional evidence that
 dynamical 3--branes are described
in terms of conformally self--dual gravity coupled to matter
-- analogous perhaps to the description of dynamical one--branes (strings)
in terms of $2d$ gravity coupled to matter.\footnote{For possibly
related suggestions about the existence of such an analogy see \cite{bon}.}

It remains to compare the
coefficients $q$ in (\ref{tim}) and $Q$ in
(\ref{tom}), as well as the beta functions. This is left for future work.
It is presently not even clear to the author whether one should {\it expect}
quantitative agreement, since the calculation leading to
(\ref{tim}) is done completely within classical string theory.
A fascinating possibility would be that studying
the flow on dynamical 3-branes might yield nonperturbative information
about string theory, instead of reproducing the classical result. 

\subsubsection*{Acknowledgements}

I thank I. Klebanov, Y. Oz, A. Tseytlin and
 E. and H. Verlinde for remarks and discussions on the ADS/CFT correspondence.
This work is supported by a Heisenberg fellowship of the DFG.

\subsection*{Appendix A: Supergravity approximation}\scs{5}

This appendix is based on ref. \cite{st}, generalizing the procedure
used there to the case at hand, which includes a Ramond-Ramond background.
We start with the superstring effective action \cite{frts}, setting $\alpha'=2$
and keeping only the metric, the dilaton and the RR-5-form for simplicity:
\begin{eqnarray}
   S^{(10)}=\int d^Dx{\sqrt G}\ e^{-2\Phi}\{ {1\over3}[10-C^{(D)}(\vec x)] -
{1\over 2\cdot5!}
e^{2\Phi}F_{mnopq}F^{mnopq}\} \  ,
\label{C5}\end{eqnarray}
where $D$ is the dimension of embedding space and the function
\ba
 C^{(D)}(\vec x)&=&D-{3}
[R-4(\nabla\phi)^2+4\Box\phi] \label{C4}
\end{eqnarray}
becomes  $\vec x$--independent and equal to the central charge when  the sigma
model  represents   a  conformal theory \cite{CP}. $C$ is defined in terms
of the variation of $S$ with respect to the constant mode of the dilaton,
and therefore does not
explicitly involve
the RR background $F^{(5)}$.
For the metric we make the ansatz of section 2:
$$ds^2=d\phi^2+e^{2\alpha(\phi)}dx_{||}^2+L^2\hat g_{mn}d\theta^md\theta^n, $$
with
$$\det\hat g\ =\ 1\ ,\ \ \ L^2\equiv e^{2\beta}.$$
The RR 5-form field strength is
$$
{1\over5!}F^2\ =\ {N^2e^{-10\beta}}\ .
$$
The string equations of motion are the reqirements that the beta
functions of the $2d$ sigma model with $10d$ target space are zero.
The beta functions for the metric and the dilaton are, respectively
(with $A,B\in\{0,...,9\}$):
\ba
0\ =\ \beta_{AB} &=&  
2\{
R_{AB}+2\nabla_{A}\nabla_{B}\Phi-{5\over 2\cdot 5!}e^{2\Phi}F_{A....}
F_B^{....}\} \ .\la{mia1}\\
0\ =\ {1\over3}({10-C)} &=&
R-4(\nabla\Phi)^2+4\Box\Phi\ .
\ea
We now make
a 1+4+5 split of the 10 coordinates into $(\phi,x^\mu,\theta^m)$,
assume that all fields are independent of $x^\mu$,
and find to leading order:
\ba
\beta_{mn}^{(10)} &=& \beta_{mn}^{(5)}\ -\ (g_{mn}''-\varphi'g_{mn}'-g^{..}g_{m.}'g_{n.}')\\
{1\over3}[10-C^{(10)}] &=&{1\over3}[5- C^{(5)}]\ +\ \varphi''\ -\ (\varphi')^2
\ ,\la{maxi}\ea
and so on, where primes denote derivatives with respect to $\phi$,
$$\varphi\ =\ 2\Phi-\log{\sqrt g}\ =\ 2\Phi\ -\ 4\alpha\ -\ 5\beta$$
is the shifted dilaton, and $\beta^{(5)}$ are the beta functions
of the $2d$ sigma model with $5d$ target space metric $g=L^2\hat g$. 

In this way, we learn from the $\beta_{\phi\phi},\beta_{\mu\nu},\beta_{mn}$ and $\Phi$--equations, respectively:
\ba
-\varphi''+4(\alpha')^2+5(\beta')^2+{1\over4}(\hat g')^2&=&{5\over8}{N^2e^{2\Phi-10\beta}}
\la{mia3}\\
\alpha'' -\varphi'\alpha'&=& {5\over8}{N^2e^{2\Phi-10\beta}}\\
\beta''- \varphi'\beta'&=& {1\over5} R-{1\over2}{N^2e^{2\Phi-10\beta}}\\
\hat g_{mn}''-\varphi'\hat g_{mn}'-\hat g^{kl}\hat g_{mk}'\hat g_{nl}'
&=&2e^{-2\beta} (\hat R_{mn}-{1\over5}\hat g_{mn}\hat R)\\
-\varphi''+(\varphi')^2&=&{{1\over3}(5- C^{(5)})}\ =\ R\ \la{mia2}\\
\Phi''-\varphi'\Phi'&=&0 \la{marcel}.
\ea
A combination of the first and the second-last equations is first
order in derivatives, constraining initial values.
The last equation follows from the other ones. The $\beta_{ij}$
equation has been split into its trace and its trace-free part.
$\hat g,\Phi$ correspond to coupling constants $\vec\lambda$,
as defined in section 2, while the other equations
(which are not independent) fix $\alpha$ and $\beta$.

E.g., $\alpha$ and $\beta$ are easily fixed {\it at} fixed points, where
$$\beta'=\hat g'=\Phi'=\varphi''=0.$$
Denoting
$$\kappa\ =\ e^\Phi\ ,\ \ Q\ =\ -\varphi'\ ,$$
so that $\alpha={Q\over4}\phi$ at the fixed point,
we read off (using $R=\hat R/L^2$)
$$\hat R\ =\ {5\over2}{N^2\kappa^2\over L^8
}\ =\ 16(\alpha')^2\ L^2\ =\ \hbox{constant.}$$
Defining $\lambda$ as in section 2 by
$$\hat R_{mn}\ =\ \lambda\ \hat g_{mn}\ \ \ \rightarrow\ \ \ \hat R\ =\ 5\lambda\ ,$$
we have
$$L^8\ =\ {\kappa^2N^2\over2\lambda}\ \ ,\ \ \ \ Q^2\ =\ {5\lambda\over L^2}\ .$$

In order to expand in the vicinity of fixed points, it is useful to
first {\it average} \cite{st}, i.e. to split the shifted dilaton
$\varphi(\vec \theta,\phi)$ into an $\vec \theta$--dependent part $\tilde\varphi(\vec\theta,\phi)$
and an $\vec \theta$--independent part $\varphi_0(\phi)$ as follows:
\be
\label{d1}
 \varphi_0(\phi)\equiv -  \log [\int d^Nx\ e^{-\varphi(\vec\theta,\phi)}]\ ,\ \ \  \
   \tilde\varphi(\vec\theta,\phi)\equiv\varphi(\vec\theta,\phi)-\varphi_0(\phi)\ .
\ee
Let us   define the  space average of a function $f(\vec \theta)$ by
\begin{eqnarray}
  <f(\vec \theta)> \equiv {{\int d^5\theta\ f(\vec \theta)\ e^{-\varphi(\vec 
\theta)}}\over
  {\int d^5\theta\ e^{-\varphi(\vec \theta)}}} \ .
\label{E1}\end{eqnarray}
Then define
\begin{eqnarray}
Q(\phi)\equiv-\varphi_0'(\phi) = - <\varphi'(\vec \theta,\phi)> \ .
\label{EE1}\end{eqnarray}
Integrating (\ref{maxi}) weighted by $e^{-\varphi}$  yields
\begin{eqnarray}
  Q'+Q^2= {1\over3} (5-\bar c),
\label{A2}\end{eqnarray}
where (\ref{E1}) has been used and we have defined  the function
\begin{eqnarray}
\bar c (\phi) = \bar c(G,B,\varphi) = <C^{(5)}(\vec \theta,\phi)>\ .
\label{AA2}\end{eqnarray}
 $\bar c$
can be considered as a generalisation of the `$c$--function' \cite{zam}
of $2d$ field theory. 

The expansion near fixed points follows as in \cite{st}.
We collectively denote $\delta \hat g, \delta\Phi, ...$ by $\vec\lambda$
as in section 2, noting i.p. that $g=L^2\hat g$.
Next we expand the above string equations of motion order by order in
$\vec\lambda$. It is easy to see that to lowest order it suffices to
set $L$ and $Q$ to their fixed point values, to ignore the terms
of order $({\vec\lambda}')^2$ in the flow equations and to set
the beta functions for the sigma model with target space metric $\hat g$
equal to those for the sigma model with metric $g$. One then obtains
 \ba L^2(\vec\lambda''\ +\ Q\vec\lambda')&=&\vec\beta^{(5)}\\
 Q^2&=& {5-\bar c\over3}\ .\ea
 The second equation is the equation of motion for the constant part of the
 shifted dilaton $\varphi$.
To next order, $Q^2$ gets a new contribution \cite{st}:
\ba
Q^2\ =\  {5-\bar c(\vec\lambda)\over3} \ +\ {1\over4}(\vec\lambda')^2\ .
\la{ganz}\ea

Finally, we define RG time near the fixed point
(where $\alpha={Q\over4}\phi$) as in section 2:
$$\tau\ \equiv\ {Q\over4}{\phi}\ \ \ \hbox{with}\ \ \ Q\equiv-{q\over L}\ .$$
We denote derivatives with respect to $\tau$ by ``\ $\dot{}$\ '', and
pull out factors of ${1\over L}$ using
$$\vec\lambda'={Q\over4}\dot{\vec\lambda}\ ,\ \ \ 
 {5-\bar c\over3}\ =\ R={1\over L^2}\hat R\ 
=\ {1\over L^2}{5-c_2\over3}\ .$$
$c_2$ is (a generalization of) the c-function
for the sigma model with embedding space metric $\hat g$ (rather than $g=L^2\hat g$).
This yields to lowest order:
\begin{eqnarray}
  {q^2\over26}(\ddot{\vec\lambda}+{4}\dot{\vec\lambda}) &=&
      \vec\beta^{(5)}\la{markus}\\ \hbox{with}\ \ 
 q^2 &=& {5-c_2\over3}\ +\ h.o.
\end{eqnarray}

\subsection*{\bf Appendix B: Conformally self-dual gravity}\scs{6}

We consider four--dimensional
Euclidean ``gravity'' with action
$$S=\int_M d^4x\ {\sqrt g}\{\lambda+\gamma R + \eta R^2 + \rho
W^2\}$$
in the limit $$\lambda=\gamma=\eta=0\ ,\ \rho\rightarrow\infty.$$
$W$ is
the Weyl tensor, the traceless part of the Riemann tensor:
$$W_{\mu\nu\sigma\tau}=R_{\mu\nu\sigma\tau}
-{1\over 2}(g_{\mu\sigma}R_{\nu\tau}+g_{\nu\tau}R_{\mu\sigma}
-g_{\mu\tau}R_{\nu\sigma}-g_{\nu\sigma}R_{\mu\tau})
+{1\over6}(g_{\mu\sigma}g_{\nu\tau}-g_{\mu\tau}g_{\nu\sigma})R.$$
In the limit $\rho\rightarrow\infty$ this theory becomes free,
as will become clear below;
this limit corresponds to an UV fixed point,
as shown by Fradkin and Tseytlin \cite{ft3}.\footnote{Related
work is \cite{fourth}.}
What makes the theory in this limit nontrivial is the conformal
anomaly term in the effective action, 
as we will review for the case of the bosonic theory.

There are two topological invariants in $4d$,
the Euler characteristic $\chi$ and the signature $\tau$ of a manifold
$$\tau= {1\over {48\pi^2}}\int d^4 x {\sqrt g} (W_+^2 - W_-^2)
\qquad\hbox{and}\qquad \chi={1\over{32\pi^2}}\int d^4 x {\sqrt g} G.
$$
with Gauss-Bonnet density
$$G=R^{\mu\nu\sigma\tau}R_{\mu\nu\sigma\tau}-4R^{\mu\nu}R_{\mu\nu}+R^2 $$
and (anti-) self-dual part of the Weyl tensor
$$W_{\pm\ \mu\nu\sigma\tau}\equiv {1\over 2}
( W_{\mu\nu\sigma\tau} \pm
 {1\over 2}\epsilon_{\mu\nu}^{\ \ \ \alpha\beta} W_{\alpha\beta\sigma\tau}).$$
{}
So after subtracting a topological invariant $\propto\tau$, we end up studying the action \cite{schm4d}
$$\rho\int d^4x{\sqrt g}W_+^2\ \ ,\ \ \ \ \rho\rightarrow\infty\ .$$

In the limit $\rho\rightarrow\infty$, all metrics are strongly suppressed in
the path integral $\int{\cal D}g\ \exp\{-S\}$
except for conformally self-dual metrics, i.e., metrics with $W_+=0$.
$W_+$ has 5 independent components, so the condition $W_+=0$
kills 5 of the 10 components of the metric.
This condition is conformally and diffeomorphism invariant. 
So the 5 remaining components of
the metric  must be the 4 diffeomorphisms and
 the $4d$ conformal
factor. In addition, there is a moduli space
of conformally self-dual metrics.
This is a $4d$ analog of the moduli space of Riemann surfaces
of $2d$ gravity.
In the simplest case of $S^4$ topology, there is no moduli space:
conformally self-dual metrics
are conformally flat,
$$\hat g = \delta e^\phi $$
up to diffeomorphisms $\delta\xi_\mu$. (For K3, e.g., the moduli space
is 57-dimensional.)
We can now split up  fluctuations around conformally self--dual metrics as
$$\delta g_{\mu\nu} = \hat g_{\mu\nu}\delta \phi
+ \nabla_{(\mu}\delta\xi_{\nu)}+\delta \bar h_{\mu\nu}.$$
The limit $\rho\rightarrow\infty$ can be regarded as the ``classical limit''
for the five other components
$\bar h_{\mu\nu}$ only, in the sense that the path integral over them 
becomes Gaussian and leaves us with a determinant
$$\det(O^\dagger O)^{-{1\over 2}}\eqno(2.4) $$
where $O^\dagger$ is the linearized $W_+$-term,
$O$ is its adjoint and $O^\dagger O$ is a $4^{th}$
order, conformally invariant,
linear differential operator \cite{ant}.
What remains is a path integral over the conformal factor
and an ordinary integral over the moduli space mentioned above.
Changing variables from $g$ to $\phi$
and $\xi$ and integrating over $\xi$ contributes, as in $2d$, 
another Jacobian
$$(\det L^\dagger L)^{1\over 2}.$$

The determinants $\det(O^\dagger O), \det (L^\dagger L)$, as well
as any conformally coupled matter fields contribute to the conformal
anomaly, analogously to $2d$.
For conformally invariant differential operators $X$:
$$\det X_{\hat g e^\phi} = \det X_{\hat g} e^{-S_i[\hat g,\phi]}$$
where the induced action $S_i$ is obtained
by integrating the trace anomaly of the stress tensor 
$$-2{{\delta S_i[\hat g,\phi]} \over{ \delta \phi}} ={\sqrt g}
<T^\mu _\mu> = {1\over{16\pi^2}}{\sqrt g}(aW^2 + bG) .$$
(There is also a term ${2\over3}a\Box R$, but it vanishes for Einstein manifolds.)
Following \cite{ft2,ant}, the conformal anomaly can be integrated:
the $4d$ analog of the (free part of the) $2d$ Liouville action is, essentially,
$$S_i[\hat g,\phi] =
 -{1\over{32\pi^2}}\int d^4 x {\sqrt{ \hat g}}\{\ 
b\ (\phi \Box^2 \phi+\hat G\phi)\ +\ a\ {\hat W}^2\phi\ \}\ .$$
Note that, after restricting to conformally self--dual metrics and
going to conformal gauge, the originally nonlocal conformal anomaly term
has become local.
Some coefficients $a$ and $b$ are according to Fradkin and Tseytlin \cite{ft1}
(there, $a={\beta\over2}, b=\beta_1-{\beta\over2},$):

\leftline{\hfill $120\ a$\hskip1.5cm $-360\ b$\hskip2cm}
\vskip 1mm\leftline{\hskip5mm conf. coupled real
scalars $\phi$ ($\triangle \sim\Box -{1\over 6}R$):\hfill
1\hskip 2.3cm 1\hskip 2.5cm}
\leftline{\hskip.5cm spin ${1\over 2}$ (two-component) fermions $\chi$:\hfill
3\hskip 2.2cm ${11\over2}$\hskip 2.4cm}
\leftline{\hskip .5cm massless gauge fields $A_\mu$:\hfill
12\hskip 2.1cm 62\hskip 2.4cm}
\leftline{\hskip.5cm Gravitino $\psi_\mu$:\hfill
$-298$\hskip 1.7cm $-548$\hskip 2.25cm}
\leftline{\hskip.5cm Graviton $(\det O^\dagger O)^{-1/2}(\det L^\dagger L)^{1/2}$:\hfill
796\hskip 1.7cm 1566\hskip 2.25cm}
\leftline{\hskip.5cm Scalar $\varphi$ with kin. term $\Box^2$, min. coupled:\hfill
--8\hskip 1.9cm --28\hskip 2.4cm}
\leftline{\hskip.5cm Fermion $\lambda$, superpartner of this $\varphi$:\hfill
$-1$\hskip 1.9cm $-{27\over2}$\hskip 2.4cm}
\vskip 5mm

Similarly as in $2d$ \cite{ddk}, we now add the conformal factor $\phi$ as an independent
new matter field \cite{ant} with standard measure
and action
$$
 {1\over{32\pi^2}}\int d^4 x {\sqrt{ \hat g}}\{\ 
 (\phi \Box^2 \phi\ -\ {1\over2}Q\ \hat G\phi\ -\ {1\over2}P\ {\hat W}^2\phi\ \}\ $$
($P$ will play a minor role here)
to the theory and demand
conformal invariance of the total theory.
This means: i), conformal anomalies have to cancel;
ii), scaling operators are dressed to have dimension 4, etc.

As for i), the free theory for $\phi$
has conformal
anomaly $$a=a_{\Box^2}+{1\over4}PQ,\hskip 1in b=b_{\Box^2}+{1\over4}Q^2,$$
where $a_{\Box^2},b_{\Box^2} $ are read off from the table.
Therefore cancellation of conformal anomalies fixes $Q,P$:
$b+b_{O^\dagger O}+b_{L^\dagger L}+b_{matter}=0,
a+a_{O^\dagger O}+a_{L^\dagger L}+a_{matter}=0$ implies
$$Q^2={1\over{90}}(N_0+11N_{1\over2}+62N_1+1538),\hskip .4in
PQ=-{1\over {30}}(N_0+6N_{1\over2}+12N_1+788).
$$
where $N_0,N_{1\over2},N_1$ are the number of conformally coupled
scalars, spin ${1\over2}$ fermions and massless gauge fields.

A key point, noted by Antoniadis and Mottola \cite{ant}, is that
due to the $\Box^2$ kinetic term for $\phi$, its propagator is logarithmic
and therefore the exponential
$e^{2\alpha\phi}$ is a scaling operator, as in $2d$
(this is very unusual in $4d$).
Its dimension comes out to be
$$\dim(e^{\alpha\phi})=-\alpha(\alpha+Q)$$
As in two
dimensions, if $\Phi_i$ is a scaling operator
of the matter theory with conformal dimension $\Delta_i$, the operator
$$O_i \equiv \int d^4 x \sqrt {\hat g} e^{\gamma_i \phi} \Phi_i$$
with $\gamma_i$ determined by
$$\Delta_i-\gamma_i(\gamma_i+Q)=4$$
is then a scale invariant operator that can be added to the action, at least
infinitesimally.

We finally mention that there
is also an analog of the $c=1$ barrier of $2d$ gravity: instead of $c\le1$ 
we have \cite{ant}
$$+360\ b\ \le\ 98.$$
This is usually satisfied, as $b$ is negative for
conventional matter.
One can also compute a scaling law e.g. for the fixed volume partition function
\cite{schm4d}:
$$Z(V)\sim V^{ -3.67...}$$
See part III of the author's thesis for details; there, $Q\equiv-{\sqrt{-4B}},\phi\rightarrow
-2{\phi\over Q},\gamma\rightarrow -{Q\over2}\gamma$.

Let us also quote the
anomaly coefficients for some supersymmetry multiplets from \cite{ft1}:

\leftline{\hfill $24\ a$\hskip1.5cm $-24\ b$\hskip2cm}
\vskip 1mm\leftline{\hskip5mm N=1 supergravity multiplet:\hfill
102\hskip 1.8cm 72\hskip 2.5cm}
\leftline{\hskip5mm N=1 scalar multiplet:\hfill
1\hskip 2.3cm ${1\over2}$\hskip 2.5cm}
\leftline{\hskip5mm N=1 vector multiplet:\hfill
3\hskip 2.3cm ${9\over2}$\hskip 2.5cm}
\leftline{\hskip5mm N=1 $\Box^2$ multiplet:\hfill
$-3$\hskip 1.8cm $-{9\over2}$\hskip 2.5cm}
\leftline{\hskip5mm N=2 supergravity multiplet:\hfill
52\hskip 2cm 41\hskip 2.5cm}
\leftline{\hskip5mm N=2 hypermultiplet:\hfill
2\hskip 2.3cm ${1}$\hskip 2.5cm}
\leftline{\hskip5mm N=2 vector multiplet:\hfill
4\hskip 2.3cm ${5}$\hskip 2.5cm}
\leftline{\hskip5mm N=2 $\Box^2$ multiplet:\hfill
2\hskip 1.7cm $-{11}$\hskip 2.5cm}
\leftline{\hskip5mm N=4 supergravity multiplet, min. coupled $\varphi$:\hfill
$-24$\hskip 1.7cm $-24$\hskip 2.5cm}
\leftline{\hskip5mm N=4 vector multiplet:\hfill
6\hskip 2.3cm $6$\hskip 2.5cm}

\end{document}